\documentclass[usenatbib,onecolumn]{mnras}
\usepackage{amsmath}
\usepackage{epsfig,color}
\usepackage{dcolumn}
\usepackage[utf8]{inputenc}
\usepackage{graphics,graphicx}
\usepackage{epstopdf}
\usepackage{float}

\graphicspath{ {eps/} }

\usepackage{url}
\usepackage{color}

\usepackage{braket}
\usepackage{lscape}

\newcommand{\cm}{cm$^{-1}$}

\newcommand{\ai}{\textit{ab initio}}
\newcommand{\Ai}{\textit{Ab initio}}

\newcommand{\duo}{{\sc Duo}}
\newcommand{\Duo}{{\sc Duo}}

\title[ExoMol line lists XXIV: SiH]{ExoMol line lists XXIV: A new hot line list
for silicon monohydride, SiH}
\date{\today}

\author[Yurchenko et al]{\large Sergei N. Yurchenko$^{1}$, Frances Sinden$^{1}$,
Lorenzo Lodi$^{1}$, Christian Hill$^{1}$, Maire N. Gorman$^{2}$, and Jonathan
Tennyson$^{1}$  \\
$^{1}$Department of Physics and Astronomy, University College London, Gower
Street, London, UK, WC1E 6BT  \\
$^{2}$ Department of Physics, Aberystwyth University, Penglais, Aberystwyth,
Ceredigion, UK, SY23 3BZ}
\date{Accepted XXXX. Received XXXX; in original form XXXX}
\pagerange{\pageref{firstpage}--\pageref{lastpage}} \pubyear{2017}

\begin{document}

\label{firstpage}

\maketitle

\begin{abstract}
SiH has long been observed in the spectrum of our Sun and other cool stars.
Computed line lists for the main isotopologues of silicon monohydride,
$^{28}$SiH, $^{29}$SiH, $^{30}$SiH and $^{28}$SiD are presented. These line
lists consider rotation-vibration transitions within the ground $X$~$^{2}\Pi$
electronic state as well as transitions to the low-lying
$A$~$^{2}\Delta$  and $a$~$^{4}\Sigma^-$ states. \Ai\ potential energy
(PECs) and dipole moment curves (DMCs) along with spin-orbit and
electronic-angular-momentum couplings between them are calculated using the
MRCI level of theory with the MOLPRO package. The PEC for the ground
X~$^2\Pi$ state is refined to available experimental data with a
typical accuracy of around 0.01~\cm\ or better. The $^{28}$SiH  line list includes
11,785
rovibronic states and 1,724,841 transitions with associated Einstein-A coefficients for angular momentum $J$ up to
$82.5$ and covering wavenumbers up to 31340 \cm\ ($\lambda$ $<$
0.319 $\mu$m). Spectra are simulated using the new line list and comparisons
made with various experimental spectra. These line lists are applicable up to
temperatures of ~5000 K, making them relevant to astrophysical objects such
as exoplanetary atmospheres and cool stars and opening up the possibility of
detection in the interstellar medium. These line lists are available at the ExoMol
(\url{www.exomol.com}) and CDS database websites.

\end{abstract}


\section{Introduction}

Silicon hydride (SiH) is a free radical formed from the cosmically abundant
elements of hydrogen and silicon, and is the simplest of the four possible
silicon hydrides. Following the first experimental measurement of the
$A$~$^{2}\Delta$ -- $X$~$^{2}\Pi$ system of SiH by \citet{30Jackson.SiH}, SiH was
observed by \citet{33Pearse.SiH} and \citet{45Babcock.SiH} in sunspots and in
the spectrum of the solar disk spectrum by \citet{64Schade.SiH}  and
\citet{66Moore.SiH}. These identifications were then corroborated by
\citet{69Sauval.SiH} using the coincidence method and also by
\citet{70LaMaxx.SiH}, who derived oscillator strengths and isotope shifts for
$^{29}$SiH and $^{30}$SiH. Furthermore, \citet{70GrSaxx.SiH} observed SiH in the
photospheric region of the sun and also subsequently calculated oscillator
strengths \citep{71GrSaxx.SiH}.

SiH has been observed in late-type stars by \citet{40Davis.SiH} and  in the
emission spectra of M- and S-type Mira variable stars \citep{55Merrill.SiH}. SiH
is also important for the modelling of M-dwarf atmospheres
\citep{95AlHaxx.SiH}, although of less importance than species with more
pronounced spectroscopic features. In exoplanetary and brown dwarf atmospheres
\citet{10ViLoFe.SiH} make the prediction that SiO should be the most abundant
silicon species in low pressure environments, with silane (SiH$_{4}$) the
most abundant at high pressures. They also predict that  SiH may
be present due to equilibrium reactions of SiH with both SiH$_{4}$ and SiO in
the presence of H$_{2}$O.

In the interstellar medium (ISM),
given the observations of similar molecules such as SiO, CH and OH,
the presence of SiH has been suggested by numerous authors
\citep{71WiPeJe.SiH,63WeBaMe.SiH,73RyElIr.SiH,74Lovas.SiH,77TuDaxx.SiH,
78DeSixx.SiH}. In particular, \citet{89HeMiWl.SiH} suggested that the abundances of both
silicon and hydrogen make SiH a likely candidate to be found in interstellar
clouds, where it so far remains undetected.

Several theoretical studies exploring the electronic and thermodynamic
properties of SiH have been made, often as parts of larger
studies of silicon hydrides. The first \ai\ electronic
structure calculations were performed by \citet{67CaHuxx.SiH} who produced a ground state
potential energy curve (PEC) using the Hartree-Fock method and
Slater-type orbital basis functions. \citet{71RaLaxx.SiH} used the Rydberg-Klein-Rees
method as well as their own modified method to calculate the PECs for the $X$~$^{2}\Pi$ and
$A$~$^{2}\Delta$ states for SiH on the basis of the known experimental data.
Later \citet{75MeRoxx.SiH} used the coupled
electron pair approximation (CEPA) method and a Gaussian-type orbital (GTO)
basis set to examine both the PEC and DMC of the SiH ground state. This work
on the PEC for the ground state has been followed up by \citet{08ShZhSu.SiH} and
\citet{09PrLuWi.SiH} using the coupled-cluster method, with the latter of these studies
forming part
of a more general, larger study of silicon-containing hydrides. Additionally,
theoretical calculations of the ground state dipole were also
performed by several authors using various methods
\citep{75MeRoxx.SiH,87Kalcher.SiH,99AjPaxx.SiH,92PaSux1.SiH,86PeLaxx.SiH}. The
transition dipole between the $X$~$^{2}\Pi$ and $A$~$^{2}\Delta$ states for SiH was
calculated by \citet{87Larsson.SiH} using the CASSCF method. \citet{86Buenker.SiH} used
SiH as a test case for his theoretical
study on the calculation of excited molecular states.

A considerable step forward was made by \citet{02KaMaMe.SiH},
who used the multi-reference configuration interaction (MRCI) method
to compute 16 electronic states of SiH.
More recently,
\citet{13ShLiZh.SiH} produced PECs for seven bound states of SiH
and gave a new set of spectroscopic parameters using, again,
the MRCI method (aug-cc-pV6Z basis set for Si and aug-cc-pV5Z
for H) and accounting for spin-orbit coupling using the Breit-Pauli
Hamiltonian.

Spin-orbit coupling in SiH has also been specifically studied by
several authors \citep{03ChSuxx.SiH, 08SoGaHa.SiH, 82StKrxx.SiH, 13ShLiZh.SiH};
more general studies of spin-orbit coupling in diatomic molecules have been
undertaken by \citet{77BrWaxx.SiH}, \citet{08LiGaZh.SiH}, \citet{90BaLexx.SiH}
and \citet{08ChSuxx.SiH}. $\Lambda$ splitting in SiH was
calculated by \citet{75WiRixx.SiH}, complementing various
experimental measurements  \citep{76FrIrxx.SiH, 79KlLiSa.SiH, 81CoRixx.SiH}.

SiH is of considerable interest to the semiconductor industry and it is a
by-product in the production of thin films for devices such as LCDs
\citep{95JaBeWa.SiH, 80TuCaGr.SiH, 81TuCaGr.SiH, 85DrToxx.SiH}. Neutral
radicals, particularly SiH$_{3}$ provide the most efficient growth in the
chemical vapour deposition process, with SiH only occurring in comparatively
negligible quantities  \citep{86RoGaxx.SiH}. It is usually produced from
SiH$_4$ using photolysis, radio frequency discharge, or by equilibrium reaction with
fluorine \citep{02KaMaMe.SiH}. Numerous studies of SiH within silane plasmas has
been undertaken \citep{80TaHiHa.SiH, 80MaNaTa.SiH, 81KaGrxx.SiH, 84ScGrKr.SiH}
and also of thermodynamic properties such as bond strength, enthalpies and
heats of formation of silicon-containing hydrides \citep{79HuNoxx.SiH,
82McGoxx.SiH, 91SaKaxx.SiH, 92LeSaWi.SiH, 09GrDixx.SiH, 10LiLoHe.SiH}.

Experimentally, rovibrational transition wavelengths for SiH within the ground
state were measured extensively during the 1980s \citep{84BrRoxx.SiH,
84BrCuE1.SiH, 85BrCuEv.SiH, 85DaIsJo.SiH, 86BeBoCh.SiH, 87SeWeUr.SiH}. However,
the first recorded wavelengths for SiH were actually for the $A$~$^{2}\Delta$ --
$X$~$^{2}\Pi$ system starting with \citet{30Jackson.SiH}, with the most recent study
by \citet{98RaEnBe.SiH} building on earlier work \citep{36Roches.SiH,
57Douglas.SiH, 65Verma.SiH, 78SiVaxx.SiH, 79KlLiSa.SiH}. Spectra for this system
were also recorded in the gas phase as part of silane glow discharge
studies \citep{89NeSuNa.SiH, 85WaMaHa.SiH, 80PeDexx.SiH, 97StMaRa.SiH}.
Additionally, the overall absorption cross sections and electronic transition
moment for this system were determined  by \citet{79Parkxx.SiH} using a shock
tube.

Limited experimental work has been undertaken for other excited states;
\citet{71BoKlPa.SiH} recorded spectra around ~ 3250 \AA\ which they attributed
to transitions involving the $B$~$^{2}\Sigma^{-}$ and $C$~$^{2}\Sigma^{+}$ states.
The $E$~$^{2}\Sigma^{+}$ -- $X$~$^{2}\Pi$ and $D$~$^{2}\Delta$ -- $X$~$^{2}\Pi$ systems
for SiH and SiD around 1907 \AA\ and 2058 \AA\ respectively were recorded by
\citet{69HeLaMc.SiH}, following earlier, preliminary work by
\citet{65Verma.SiH}. Finally, \citet{89JoHuxx.SiH} located a state at 46700
$\pm$ 10 \cm\ which they classified as either $^{2}\Pi$ or $^{2}\Sigma^{+}$
using resonance-enhanced multiphoton ionization (REMPI) spectroscopy.

The Cologne Database for
Molecular Spectroscopy (CDMS) lists 125 purely rotational transition
lines within the vibrational ground state spanning the 0.2--5275 GHz range (radio and microwave
spectral range) \citep{cdms}.
\citet{11Kurucz.db} has compiled a larger line list of 78,286 transitions, last
updated in 1998 for $J$ up to 37.5. Oscillator strengths and Franck-Condon
factors have been measured by \citet{71SmLixx.SiH}
whilst measurements of lifetimes have been recorded
by \citet{69Smith.SiH} and \citet{84BaBeDu.SiH},
\citet{89NeSuNa.SiH}.

The ExoMol project  aims at providing line lists of
spectroscopic transitions for key molecular species which are likely to be
important in the atmospheres of extrasolar planets and cool stars
\citep{jt528,jt631}.  This is essential for the continued exploration of newly
discovered astrophysical objects such as exoplanets, for which there is an
increasing desire to characterise their atmospheric compositions.  The
methodology of the line list production for diatomics is discussed by
\cite{jt626}. ExoMol has already provided rotation-vibration line lists
for the closed shell silicon-containing molecules SiO \citep{jt563} and
SiH$_4$ \citep{jt701}. Given the astronomical interest in SiH, we present a new
line list for SiH applicable for temperatures up to 5000~K.

\section{Method}

The procedure  used here for the calculation of SiH line lists
is the established ExoMol methodology of refining \ai\ results to available
experimental data. Because of the absence of a complete set of potential energy
curves (PECs), dipole moment curves (DMCs), spin-orbit curves (SOC) and
electronic angular momentum coupling curves (EAMC) in the literature, \ai\ calculations
of the four lowest-lying electronic states were performed using the
program MOLPRO \citep{molpro.method, 12WeKnKn.methods}.

The \ai\ PEC, SOC, and EAMC curves were produced at the  MRCI  level of theory
in conjunction with the aug-cc-pwCVQZ basis sets
\citep{89Duxxxx.ai, 93WoDuxx.ai,02PeDuxx.ai} with relativistic, core-correlation
effects and Davidson correction taken into account. The PECs of
$X$~$^2\Pi$ and  $A$~$^{2}\Delta$ as well as SOCs, EAMCs between the $X$, $A$, and $B$
states were then refined using  \Duo\ \citep{jt609} and the available
experimental data.

\begin{figure}
\centering
\includegraphics[width=220pt]{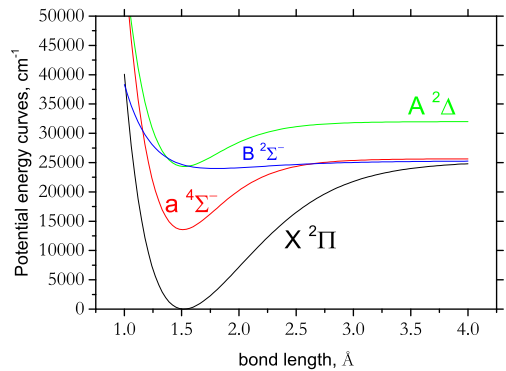}
\caption{Potential energy curves of SiH used in the line list production. The
$X$~$^{2}\Pi$ and $A$~$^{2}\Delta$ PECs have been refined, the $a$~$^{4}\Sigma^-$ is
\ai\ and the $B$~$^{2}\Sigma^-$ PEC is an artificial object used to improve the
description of the $\Lambda$-doubling in the $X$-state spectra (see section \ref{sec:refin}).}
\label{calc-pecs}
\end{figure}



\section{Dipole moment curves}\label{sec:dipole}

The DMC of the $X$~$^2\Pi$ state was computed using MRCI/aug-cc-pwCV5Z-DK \citep{89Duxxxx.ai,89RaTrPo.ai}
with the core-correlation and relativistic effects, the latter using the Douglas-Kroll-Hess
method \citep{DKH.SiH} and including the Davidson correction
\citep{74LaDaxx.ai}.
The finite field method was used.
For a discussion of the calculation of
DMCs using the expectation and finite field methods see
\citet{jt475}.

\begin{figure}
\centering
\includegraphics[width=220pt]{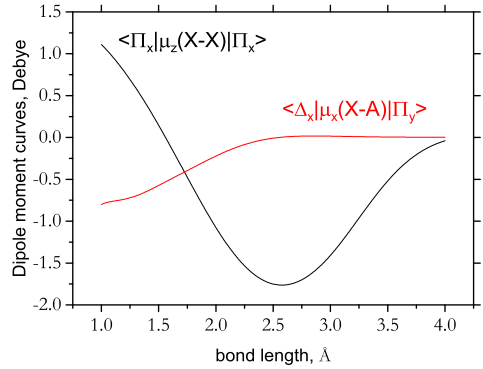}
\caption{Dipole moment curve (DMC) calculated for the $X$~$^{2}\Pi$ state and the
transition dipole moment (TDM) for the $A$~$^{2}\Delta$ -- $X$~$^{2}\Pi$ state.}
\label{f:dipoles} 
\end{figure}

The dipole moment for the $X$~$^{2}\Pi$ state and the transition dipole moment
between the $A$~$^{2}\Delta$ -- $X$~$^{2}\Pi$ state are shown in Fig.
\ref{f:dipoles}. Our equilibrium value of the $X$~$^{2}\Pi$ DMC is 0.097~D.
Previous theoretical estimates range between 0.076~D \citep{86AlScxx.SiH} and
0.173~D  \citep{99AjPaxx.SiH} with other estimates being
0.122~D	\citep{87Larsson.SiH, 92PaSux1.SiH},
0.125~D \citep{88MaMaRo.SiH},
0.117~D	\citep{86PeLaxx.SiH},
0.140~D \citep{75MeRoxx.SiH} and 0.160~D \citep{65Huzina.SiH}.

The \ai\ transition dipole moment components $\mu_\alpha$ between the
$A$~$^{2}\Delta$ and $X$~$^{2}\Pi$ states can be defined as  the matrix elements
between the Cartesian $| \Delta_{\alpha} \rangle$ and $|\Pi_{\alpha} \rangle$
components  of the corresponding electronic eigenfunctions (where
$\alpha=x,y,z$), which is also MOLPRO's basis set convention. For the `equilibrium' value
of $\langle  \Delta_z |\mu_y |\Pi_y \rangle$ (taken at $r = 1.52$~\AA) we
obtained 0.585~D. This matrix element is connected to the spherical tensor dipole
representation by
\begin{equation}
  \langle  \Lambda=2 |\mu_{+} |\Lambda = 1 \rangle =  \sqrt{2} \langle  \Delta_x
|\mu_y |\Pi_y \rangle = 0.827~{\rm D},
  \label{e:Lambda}
\end{equation}
where $\mu_+ = (-\mu_x + i\mu_y )/\sqrt{2}$. Here $|1 \rangle \equiv |^2\Pi
\rangle$  and $|2 \rangle \equiv | ^2\Delta \rangle$ are eigenfunctions of the
$\hat{L}_z$ operator and also linear combinations of $|\Pi_x\rangle,
|\Pi_y\rangle$ and $|\Delta_z\rangle, |\Delta_{xy}\rangle$, respectively.  The
tensorial representation of the dipole moment ($\mu_+$) has been generally
recommended \citep{80WhScTa}. \citet{87Larsson.SiH} obtained a CASSCF value
of $\mu_+ =  0.706$~D  using a 5-electrons in 9 orbitals $(3\sigma,2\pi,1\delta)$
active space (see Table V of cited paper)
and a basis set approximately equivalent in size to the cc-pVTZ one.
It should be noted that
the latter value better reproduces the observed $A$--$X$ lifetime (see
discussion below).
As suggested by \citet{87Larsson.SiH}, it is important to include at least one set of
(doubly degenerate) $\delta$ orbitals into CAS. We also found a strong variation of the $A$--$X$
transition dipole with respect to the active space used.
Our final choice for the active space was a rather large 5-electrons in 15 orbitals complete active space
comprising ($C_{2v}$ symmetry labels) 8~$a_1$, 3~$b_1$, 3~$b_2$ and 1~$a_2$ active orbitals; the five
core orbitals (3 of $a_1$ symmetry and one for both $b_1$ and $b_2$ symmetries)
were kept doubly occupied (i.e., excluded from the active space). CASSCF calculations used state-averaging
over the lowest $\Sigma^+$, $\Sigma^-$, $\Pi$ and $\Delta$ states. Our choice of active space was also
motivated by reason of numerical stability and convergence of the calculations (see also the discussion at the end of section
\ref{sec:linelists}).
The aug-pV(Q+d)Z basis set was used. Core-correlation and relativistic corrections had only
marginal effects and were not taken into account for this property. In any case
contributions from these effects tend to cancel \citep{jt573}.

In order to reduce the numerical noise when computing the line-strengths using
the \Duo\ program, we followed the recommendation by \citet{16MeMeSt} and
represented these two DMCs analytically. The following expansion with a
damped-coordinate was employed to represent our dipole moment functions:
\begin{equation}\label{e:dipole}
  \mu(r) = (1-\xi) \sum_{n \ge 0} d_n z^n   +d_{\infty} \, \xi ,
\end{equation}
where $\xi$ is the \v{S}urkus variable \citep{84SuRaBo.method}
\begin{equation}
\label{e:surkus}
\xi=\frac{r^{p}-r^{p}_{\mathrm{ref}}}{r^{p}+r^{p}_{\rm ref }}
\end{equation}
with $p$ as a parameter, and $r_{\rm ref }$ as a reference position
and $z$ is given by
\begin{equation}
z = (r-r_{\rm ref})\, e^{-\beta_2 (r-r_{\rm ref})^2-\beta_4 (r - r_{\rm
ref})^4}.
\end{equation}
The expansion parameters $d_n$,  $d_{\infty}$ (the value of the dipole at $r\to \infty$),  $\beta_2$ and $\beta_4$ (damping factors)  are collected in Table~\ref{t:dip} and are given in supplementary material as part of the \Duo\ input file, while the functional form has been implemented into \Duo.


\begin{table}
\centering
\caption{Expansion parameters of the dipole moment functions, $X$--$X$ $\langle
\Pi_x |\mu_z |\Pi_x \rangle$ and $X$--$A$ $\langle  \Delta_x |\mu_y |\Pi_y
\rangle$, see Eq.~(\protect\ref{e:dipole}). The units are \AA\ and Debye.}
\begin{tabular}{rrr}
\hline
\hline
    Parameter       &    \multicolumn{1}{c}{$X$--$X$}                 &    \multicolumn{1}{c}{$X$--$A$}            \\
\hline
 $  r_{\rm ref}    $&$                   1.5202  $&$            1.5202  $\\
 $  \beta_2        $&$                    0.216  $&$             0.460  $\\
 $  \beta_4        $&$               0.05979580  $&$              0.01  $\\
 $  p              $&$                        1  $&$                 8  $\\
 $  d_0            $&$             0.0970036013  $&$      0.4857112345  $\\
 $  d_1            $&$            -2.3766347830  $&$      0.5013535548  $\\
 $  d_2            $&$            -1.3617433491  $&$      0.7190055685  $\\
 $  d_3            $&$             0.1146612097  $&$      0.1181888386  $\\
 $  d_4            $&$             0.1371530465  $&$     -2.2686306105  $\\
 $  d_5            $&$             0.2436343227  $&$     -2.2419525352  $\\
 $  d_6            $&$                           $&$      2.3729868957  $\\
 $  d_{\infty}       $&$                        0  $&$                 0  $\\
\hline
\hline
\end{tabular}
\label{t:dip}
\end{table}

The PECs shown in Fig.~\ref{calc-pecs} are those calculated in this work with
the $X$~$^{2}\Pi$ and  $A$~$^{2}\Delta$ states having been refined, $a$~$^{4}\Sigma^-$ is
\ai\, and $B$~$^{2}\Sigma^-$ is a reference curve used to improve the description
of the $\Lambda$-doubling (see discussion below). Qualitatively, in terms of
general behaviour these compare favourably to \citet{02KaMaMe.SiH} who used
larger basis sets (aug-cc-pV5Z and aug-cc-pV6Z) to calculate a number of PECs
and SOCs for the low-lying electronic states of SiH.


\section{Refinement}\label{sec:refin}

In order to refine our model for the $X$~$^{2}\Pi$  and $A$~$^{2}\Delta$ states the
experimental frequencies were collected from the papers given in Table
\ref{t:table-exp-data}. These measurements span $J$ up to 18.5 in the $(0-1)$, $(0-1)$ and $(1-1$)
bands, and $J$ up to $10.5$ in the $(2-0)$ and $(3-2)$ bands. Using the
\textsc{MARVEL} program (Measured Active Rotational-Vibrational Energy Levels)
\citep{MARVEL,12FuCsi.method}, 337  energy levels were determined from
894 transitions.

\begin{table}
\caption{Summary of Experimental Data used for refining the \ai\ $X$~$^{2}\Pi$ PEC
if SiH. CTS=Czerny-Turner spectrograph, DLAR=diode laser absorption
spectroscopy, FTS=Fourier Transform Spectrometer}
\begin{tabular}{lccccc}\hline
Study	&	Method	&	System	&	J	&	$\nu$	&	
Wavenumber Range (\cm)	\\
\hline
\hline
\citet{79KlLiSa.SiH}	&	CTS	&	A~$^{2}\Delta$-X~$^{2}\Pi$	
&	0.5 – 14.5	&	(0,0)	&	23 958 – 24 399	\\
\citet{85DaIsJo.SiH}	&	DLAR	&	X~$^{2}\Pi$ - X~$^{2}\Pi$	
&	0.5 – 9.5	&	(1, 0)	&	1838 – 2094	\\
\citet{86BeBoCh.SiH}	&	FTS	&	X~$^{2}\Pi$ - X~$^{2}\Pi$	
&	0.5 – 15.5	&	(1,0), (2, 1), (3, 2)	&	1704 – 2142	
\\
\citet{98RaEnBe.SiH}	&	FTS	&	A~$^{2}\Delta$-X~$^{2}\Pi$	
&	1.5 – 18.5	&	(0, 0), (1, 1)	&	23 644 – 24 461 	
\\
\hline
\hline
\end{tabular}
\label{t:table-exp-data}
\end{table}

The \ai\ PECs, SOCs and EAMCs were refined by fitting to these derived
MARVEL energy levels, which were then complemented with the original
experimental frequencies. We used the Extended Morse Oscillator (EMO) analytical
function to represent the PECs in the fits,  which has the form
\begin{equation}\label{e:pec}
V(r)=V_{\rm e}\;\;+\;\;(A_{\rm e} - V_{\rm
e})\left[1\;\;-\;\;\exp\left(-\sum_{k=0}^{N} B_{k}\xi^{k}(r-r_{\rm e})
\right)\right]^2,
\end{equation}
where $D_{\rm e} = A_{\rm e} - V_{\rm
e}$ is the dissociation energy, $r_{\rm e}$ is an equilibrium
distance of the PEC, and $\xi$ is the \v{S}urkus variable with $r_{\rm ref} =
r_{\rm e }$.
Note that $p$ and $N$ can have different values in the short ($r\le r_{\rm e}$) and long ($r > r_{\rm
e}$) regions, i.e. $p_{\rm s}(N_{\rm s})$ and $p_{\rm l}(N_{\rm l})$, respectively.
The parameters are given as supplementary data together with the actual PECs for convenience.

\begin{figure}
\centering
\includegraphics[width=220pt]{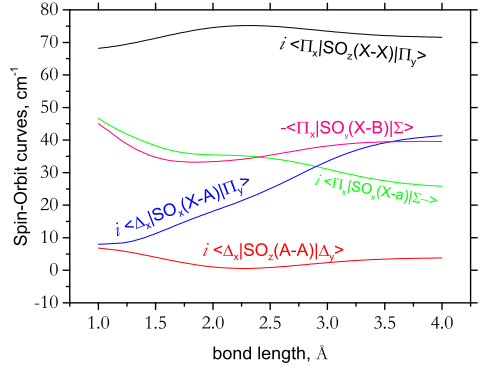}
\caption{The SO couplings (Cartesian representation as computed by MOLPRO).
The $X$--$X$, $X$--$A$, $X$--$B$ SOCs were refined.}
\label{spin-orbit} 
\end{figure}

\begin{figure}
\centering
\includegraphics[width=220pt]{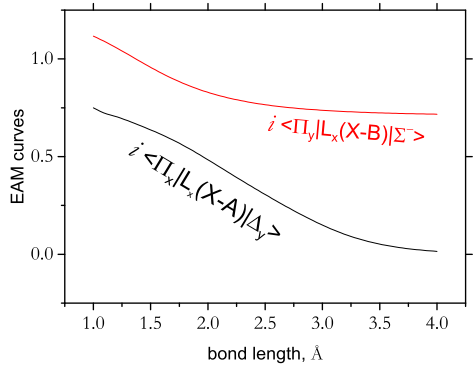}
\caption{Electronic angular momentum coupling curves (EAMC) couplings (Cartesian representation as in MOLPRO). Both the $X$--$A$
and $X$--$B$ EAMCs were refined.}
\label{f:eamc} 
\end{figure}

During the fitting of the $X$ and $A$ states, various electronic couplings involving
this electronic state were included. The refined coupling curves are shown in
Figs.~\ref{spin-orbit} and \ref{f:eamc}. To represent the $X$--$X$ and $A$--$A$ SOCs,
the expansion in Eq.~\eqref{e:dipole} was used.
The non-diagonal $X$--$a$, $X$--$A$, $X$--$B$ SOCs as well as the EAMCs were morphed using
the following  expansion in terms of the \v{S}urkus variable:
\begin{equation}
\label{e:bob}
F(\xi)= (1-\xi) \sum^{N}_{k=0}B_{k}\, \xi^{k} + \xi\, B_{\infty}.
\end{equation}
The expansion parameters are given in supplementary material as part of the \Duo\ input file.

For $\Pi$ states $\Lambda$-doubling is predominantly caused by the
interaction with $\Sigma$-states \citep{79BrMexx.methods}. In the case
of $X$~$^{2}\Pi$ of SiH, the closest $\Sigma$ state is
$B$~$^2\Sigma^-$.  This interaction is still much weaker than the SO
coupling, giving splittings of roughly 0.1 \cm\ for $J = 0.5$ (see
Table~\ref{t:obs-calc:vib}) with the two components having almost the
same intensity \citep{65Verma.SiH}. Theoretically, excited states such
as the $A$~$^{2}\Delta$ should also exhibit $\Lambda$-doubling,
although because of the small splittings this has been difficult to
observe experimentally.

The EAM coupling between the $B$~$^{2}\Sigma^{-}$ and $X$~$^{2}\Pi$ states is shown
in Fig.~\ref{f:eamc}. Since the  $B$~$^{2}\Sigma^{-}$ PEC is weakly bound and
\Duo\ currently can only work with properly bound potential energy curves, we
used a dummy curve with a shallow minimum for the $B$~$^{2}\Sigma^{-}$ PEC
as a reference object. This was sufficient to produce the correct
$\Lambda$-doubling effect for all $J$  values considered. It should be noted that
inclusion of an effective $\Lambda$-doubling curve is available in \Duo\
\citep{jt609} but did not produce a correct $J$-dependence and therefore was
discarded. In  order to model the $X$--$B$ EAM coupling the
\v{S}urkus-expansion in Eq.~\eqref{e:bob} was used.

As well as SO and EAMC, other empirical corrections such as Born-Oppenheimer
breakdown (BOB) \citep{lr07} and spin-rotation (SR) \citep{93Kato.methods} were
used to provide greater accuracy, see \citet{jt609}.
The SR within the $X$ state and the $X$--$X$ and $A$--$A$ SO couplings were
represented by the form given by Eq.~(\ref{e:dipole}) with the remaining SOC,
EAMC, SR and the BOB curves were represented by Eq.~\eqref{e:bob}.
The parameters obtained as well as all curves specifying our final model
form part of the \Duo\ inputs are provided in the supplementary
material.

The dissociation energy is poorly constrained by the low-lying vibrational state
included in the fit. Initially the $D_{\rm 0}$ value was set to the experimental
estimate of 2.98~eV ($\pm$ 0.03~eV) from \citet{87BeGrCh.SiH} and then allowed to 
float to values near this value.
Our final value for the dissociation energy $D_0$ for the $X$ state is 3.020~eV, which
corresponds to $D_{\rm e} = 3.136$~eV and the zero-point energy of 941.23~\cm\
($v=0, J=0.5, +$). 
Our final dissociation energy $D_{0}$ of the $A$ state is 0.84~eV ($D_{\rm e} =
0.95$~eV).

\begin{figure}
\centering
\includegraphics[width=320pt]{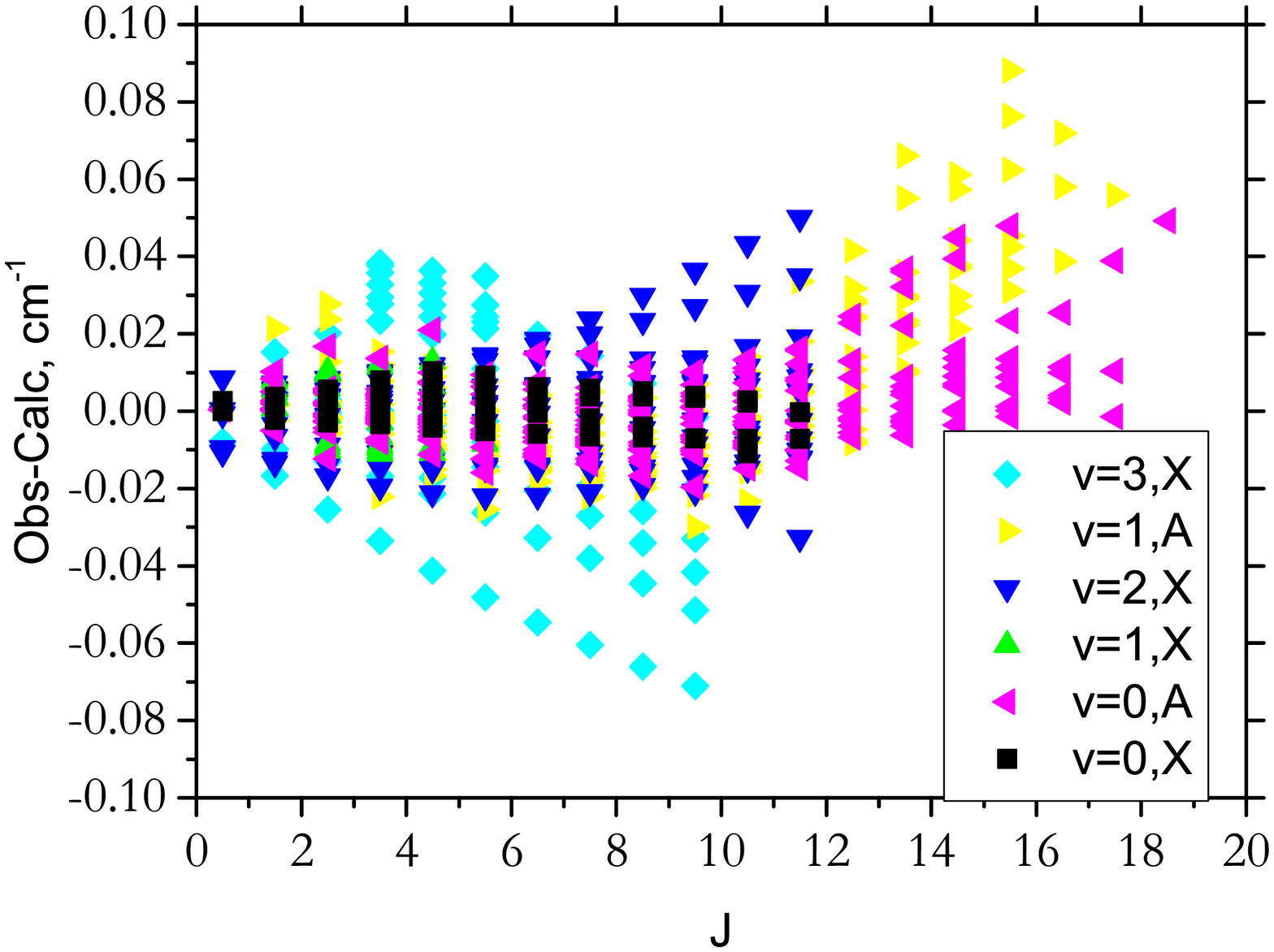}
\caption{Visual representation of difference in calculated and experimentally
measured frequencies as a function of $J$ for low-lying vibrational states of
$^{28}$SiH in its $X$~$^{2}\Pi$ and $A$~$^{2}\Delta$ electronic states.}
\label{f:obs-calc}
\end{figure}

Figure~\ref{f:obs-calc} offers a visual comparison of the Obs.-Calc. residuals
as a function of $J$, with the various vibrational and vibronic bands are
indicated. Table \ref{t:obs-calc:rot} presents a representative sample of data
for low $J$ for $v=0$ showing the comparison and table \ref{t:obs-calc:vib}
demonstrates the accuracy when the vibrational number $v$ is varied. 400 $X$--$X$
transition wavenumbers ($J'\le 16.5$) are reproduced with the  root-mean-square
(rms) error of 0.015~\cm, 51 pure rotational energies ($J\le 11.5$) collected
from CDMS are reproduced with an rms error of 0.005~\cm, 494 rovibronic $A$--$X$
transition wavenumbers ($J'\le 18.5$) are reproduced with an rms error of
0.016~\cm.  The accuracy of the pure rotational energies is comparable to that
of the typical effective Hamiltonian approaches.

\begin{table}
\caption{Example of Obs.$-$Calc. residuals, in \cm, for $^{28}$SiH $v=0$ levels illustrating the rotational accuracy
of the $X$ state of our refined model.}
\begin{tabular}{rcrrrr}
\hline
\hline
$J$ & Parity & $\Omega$ & Obs. & Calc. & Obs-Calc.  \\
\hline
 0.5 &$   -    $&$       -0.5  $&      0.1000   &      0.0972   &$     0.0028
$\\
 1.5 &$   +    $&$        0.5  $&     21.0376   &     21.0336   &$     0.0040
$\\
 1.5 &$   +    $&$        1.5  $&    151.5592   &    151.5585   &$     0.0007
$\\
 1.5 &$   -    $&$       -0.5  $&     20.8457   &     20.8469   &$    -0.0012
$\\
 1.5 &$   -    $&$       -1.5  $&    151.5515   &    151.5510   &$     0.0006
$\\
 2.5 &$   +    $&$        0.5  $&     55.6671   &     55.6698   &$    -0.0026
$\\
 2.5 &$   +    $&$        1.5  $&    190.5415   &    190.5407   &$     0.0009
$\\
 2.5 &$   -    $&$       -0.5  $&     55.9368   &     55.9318   &$     0.0050
$\\
 2.5 &$   -    $&$       -1.5  $&    190.5710   &    190.5697   &$     0.0013
$\\
 3.5 &$   +    $&$        0.5  $&    104.8464   &    104.8407   &$     0.0056
$\\
 3.5 &$   +    $&$        1.5  $&    245.0493   &    245.0471   &$     0.0021
$\\
 3.5 &$   -    $&$       -0.5  $&    104.5183   &    104.5223   &$    -0.0040
$\\
 3.5 &$   -    $&$       -1.5  $&    244.9792   &    244.9780   &$     0.0012
$\\
 4.5 &$   +    $&$        0.5  $&    167.4554   &    167.4606   &$    -0.0052
$\\
 4.5 &$   +    $&$        1.5  $&    314.7410   &    314.7398   &$     0.0013
$\\
 4.5 &$   -    $&$       -0.5  $&    167.8197   &    167.8136   &$     0.0061
$\\
 4.5 &$   -    $&$       -1.5  $&    314.8734   &    314.8703   &$     0.0031
$\\
 5.5 &$   +    $&$        0.5  $&    244.9044   &    244.8982   &$     0.0062
$\\
 5.5 &$   +    $&$        1.5  $&    399.9092   &    399.9051   &$     0.0042
$\\
 5.5 &$   -    $&$       -0.5  $&    244.5274   &    244.5337   &$    -0.0063
$\\
 5.5 &$   -    $&$       -1.5  $&    399.6918   &    399.6907   &$     0.0011
$\\
 6.5 &$   +    $&$        0.5  $&    335.7692   &    335.7763   &$    -0.0070
$\\
 6.5 &$   +    $&$        1.5  $&    499.6914   &    499.6909   &$     0.0005
$\\
 6.5 &$   -    $&$       -0.5  $&    336.1354   &    336.1292   &$     0.0062
$\\
 6.5 &$   -    $&$       -1.5  $&    500.0167   &    500.0116   &$     0.0051
$\\
 7.5 &$   +    $&$        0.5  $&    441.5298   &    441.5240   &$     0.0058
$\\
 7.5 &$   +    $&$        1.5  $&    615.0542   &    615.0480   &$     0.0061
$\\
 7.5 &$   -    $&$       -0.5  $&    441.1975   &    441.2052   &$    -0.0076
$\\
 7.5 &$   -    $&$       -1.5  $&    614.5987   &    614.5994   &$    -0.0006
$\\
 8.5 &$   +    $&$        0.5  $&    560.8100   &    560.8179   &$    -0.0079
$\\
 8.5 &$   +    $&$        1.5  $&    744.2736   &    744.2760   &$    -0.0024
$\\
 8.5 &$   -    $&$       -0.5  $&    561.0863   &    561.0811   &$     0.0052
$\\
 8.5 &$   -    $&$       -1.5  $&    744.8800   &    744.8731   &$     0.0069
$\\
 9.5 &$   +    $&$        0.5  $&    694.7840   &    694.7799   &$     0.0041
$\\
 9.5 &$   +    $&$        1.5  $&    889.3533   &    889.3460   &$     0.0074
$\\
 9.5 &$   -    $&$       -0.5  $&    694.5847   &    694.5926   &$    -0.0079
$\\
 9.5 &$   -    $&$       -1.5  $&    888.5761   &    888.5810   &$    -0.0049
$\\
10.5 &$   +    $&$        0.5  $&    842.4815   &    842.4890   &$    -0.0075
$\\
10.5 &$   +    $&$        1.5  $&   1047.3665   &    1047.374   &$    -0.0084
$\\
10.5 &$   -    $&$       -0.5  $&    842.5838   &    842.5813   &$     0.0025
$\\
10.5 &$   -    $&$       -1.5  $&   1048.3332   &    1048.325   &$     0.0076
$\\
11.5 &$   +    $&$        0.5  $&   1004.4299   &    1004.429   &$     0.0004
$\\
11.5 &$   -    $&$       -0.5  $&   1004.4433   &    1004.450   &$    -0.0070
$\\
\hline
\hline
\end{tabular}
\label{t:obs-calc:rot}
\end{table}

\begin{table}
\caption{Example of Obs.$-$Calc. residuals, in \cm, illustrating the vibrational accuracy
of our refined model. Energies for $^{28}$SiH are given relative to the $v=0$, $J=0.5$,
+, $\Omega = 0.5$ level.}
\begin{tabular}{rcccrrrr}
\hline
\hline
$J$ & Parity & State & $v$  &  $\Omega$ & Obs. & Calc. & Obs-Calc. \\
\hline
     0.5   &$     +      $&     $X$     &$      1  $&$    0.5   $&       1970.3041
  &      1970.3130   &$   -0.0089   $\\
     0.5   &$     +      $&     $X$     &$      2  $&$    0.5   $&       3869.6575
  &      3869.6600   &$   -0.0025   $\\
     0.5   &$     +      $&     $X$     &$      3  $&$    0.5   $&       5698.7697
  &      5698.7789   &$   -0.0092   $\\
     0.5   &$     -      $&     $X$     &$      0  $&$   -0.5   $&          0.0859
  &         0.0972   &$   -0.0113   $\\
     0.5   &$     -      $&     $X$     &$      1  $&$   -0.5   $&       1970.4103
  &      1970.4108   &$   -0.0005   $\\
     0.5   &$     -      $&     $X$     &$      2  $&$   -0.5   $&       3869.7405
  &      3869.7585   &$   -0.0180   $\\
     0.5   &$     -      $&     $X$     &$      3  $&$   -0.5   $&       5698.8651
  &      5698.8763   &$   -0.0112   $\\
     1.5   &$     +      $&     $X$     &$      0  $&$    0.5   $&         21.0376
  &        21.0336   &$    0.0040   $\\
     1.5   &$     +      $&     $X$     &$      0  $&$    1.5   $&        151.5592
  &       151.5585   &$    0.0007   $\\
     1.5   &$     +      $&     $X$     &$      1  $&$    0.5   $&       1990.7716
  &      1990.7826   &$   -0.0110   $\\
     1.5   &$     +      $&     $X$     &$      1  $&$    1.5   $&       2122.1435
  &      2122.1638   &$   -0.0203   $\\
     1.5   &$     +      $&     $X$     &$      2  $&$    0.5   $&       3889.5532
  &      3889.5673   &$   -0.0141   $\\
     1.5   &$     +      $&     $X$     &$      2  $&$    1.5   $&       4021.7593
  &      4021.7763   &$   -0.0170   $\\
     1.5   &$     +      $&     $X$     &$      3  $&$    0.5   $&       5718.0995
  &      5718.1237   &$   -0.0242   $\\
     1.5   &$     +      $&     $A$     &$      0  $&$    1.5   $&      24268.0465
  &     24268.0614   &$   -0.0149   $\\
     1.5   &$     +      $&     $A$     &$      1  $&$    1.5   $&      25928.2856
  &     25928.2823   &$    0.0033   $\\
     1.5   &$     -      $&     $X$     &$      0  $&$   -0.5   $&         20.8283
  &        20.8469   &$   -0.0186   $\\
     1.5   &$     -      $&     $X$     &$      0  $&$   -1.5   $&        151.5420
  &       151.5510   &$   -0.0090   $\\
     1.5   &$     -      $&     $X$     &$      1  $&$   -0.5   $&       1990.5945
  &      1990.5943   &$    0.0002   $\\
     1.5   &$     -      $&     $X$     &$      1  $&$   -1.5   $&       2122.1346
  &      2122.1566   &$   -0.0220   $\\
     1.5   &$     -      $&     $X$     &$      2  $&$   -0.5   $&       3889.3696
  &      3889.3775   &$   -0.0079   $\\
     1.5   &$     -      $&     $X$     &$      2  $&$   -1.5   $&       4021.7541
  &      4021.7694   &$   -0.0153   $\\
     1.5   &$     -      $&     $X$     &$      3  $&$   -0.5   $&       5717.9357
  &      5717.9355   &$    0.0002   $\\
     1.5   &$     -      $&     $A$     &$      0  $&$   -1.5   $&      24268.0560
  &     24268.0614   &$   -0.0054   $\\
     1.5   &$     -      $&     $A$     &$      1  $&$   -1.5   $&      25928.2705
  &     25928.2823   &$   -0.0118   $\\
     \hline
     \hline
\end{tabular}
\label{t:obs-calc:vib}
\end{table}

\section{Line lists}\label{sec:linelists}

\Duo\ was  used to solve the fully coupled Schr\"{o}dinger
equation for the four lowest bound electronic states of SiH using our refined
curves. The details of the \duo\ methodology used for building accurate,
empirical line lists for diatomic molecules has been extensively discussed
elsewhere \citep{jt589,jt598,jt599,jt632,jt609}. A grid-based sinc basis of 501
points spanning 1 to 4 \AA\ was used, selecting the 40, 20, 40, and 10 lowest
vibrational eigenfunctions of the $X$~$^{2}\Pi$, $a$~$^{4}\Sigma^{-}$,
$A$~$^{2}\Delta$, $B$~$^{2}\Sigma^{-}$ states respectively.

The line list produced for $^{28}$SiH contains 1,724,841 transitions and 11,785
states covering frequencies up to 31340 \cm. The line
lists for the isotopologues $^{29}$SiH, $^{30}$SiH and $^{28}$SiD were generated
using the same methodology by simply changing the nuclear masses to the
corresponding values (see Table~\ref{t:stats}). For compactness and ease of use the line lists are
separated into energy state and transitions files using the standard ExoMol
format \citep{jt631}. Tables \ref{table:states} and \ref{table:trans} show
extracts from the States and Transition files, respectively.  The full line
lists for all isotopologues considered can be downloaded from
\url{www.exomol.com} and from the CDS database.

\begin{table}
\caption{Statistics for line lists for all four isotopologues of SiH.
}
\begin{tabular}{lrrrr}
\hline\hline
& $^{28}$SiH & $^{29}$SiH  & $^{30}$SiH  & $^{28}$SiD   \\
\hline
$J_{\rm max}$          	& 82.5      	& 82.5 		& 82.5 		& 113.5       \\
$\nu_{\rm max}$ (\cm)  	& 31337.3     	& 31337.3 	&  31337.3 	& 31337.3   \\
$E'_{\rm max}$ (\cm)   	& 31337.3     	&   31337.3 	& 31337.3 	& 31337.3      \\
$E''_{\rm max}$ (\cm)   & 31337.3    	&  31337.3 	& 31337.3 	& 31337.3       \\
number of energies     	& 11785		& 11796		& 11808		& 21230	\\
number of lines        	& 1724841	& 1726584	& 1728386	& 3520657	\\
\hline
\end{tabular}
\label{t:stats}
\end{table}

As part of the ExoMol line lists, we now provide lifetimes and Land\'{e} $g$-factors for the states
involved \citep{jt624}. The methodology used to compute them is detailed in \citet{jt656}.
The lifetime of the $A$ state ($v=0$, $J\le 11.5$)  was measured in
laser-induced fluorescence experiments
by \citet{84BaBeDu.SiH}   to be $534 \pm 23$ ns and by \citet{84ScGrKr.SiH} to be
$530\pm 2$~ns, see also a review of other measurements by \citet{84BaBeDu.SiH}.
Our value is 400~ns for the $J'=10.5$ ($0-0$).
\citet{87Larsson.SiH} in his CASSCF calculations showed that the $X$--$A$ transition dipole moments
depends strongly on the active space (see Table V of cited paper)
and that the closest agreement with the
experimental lifetime was not for the largest one.
Among \citet{87Larsson.SiH}'s computed lifetimes the one closest to experiment is 484~ns ($\mu_{+} = 0.706$~D)
using a 5-electrons in 9 orbitals $(3\sigma,2\pi,1\delta)$ active space.
Using a similar 5-electrons in 10 orbitals active space comprising 5~$a_1$, 2~$b_1$, 2~$b_1$ and 1~$a_2$ orbitals
we have obtained a similar value of the equilibrium dipole moment ($\mu_{+} = 0.706$~D).
However, we could not converge our
calculations of the DMC with this choice of the active space for all the geometries considered;
we managed to obtain convergence for all values of $r$ between $0.7$ and
5~\AA\ using the large 5-electrons in 15 orbitals
active space described at the end of section \ref{sec:dipole}
which, however, gave a slightly too high
value of $\mu_{+}$ (0.789~D). We therefore decided to re-scale the
(11,4,4,1) DMC by a factor $\sqrt{400/530} \approx 0.869$, thus bringing the lifetime of the $J'=10.5$
($0-0$) level to 530~ns. It should be noted that the lifetime of the $v=0$ state gradually increases to ~$800$~ns at $J=32.5$.



\begin{table*}
\centering
\caption{Extract from the states file of the $^{28}$Si$^{1}$H line list.
}
\label{table:states}
{\tt
 \begin{tabular}{rrrrrrcclrrrr}
 \hline
 \hline
$n$	&	Energy (\cm)	&	$g_i$	&	$J$	& $\tau$ &	$g$-factor	&	Parity	&	e/f	&	State	&	$v$	&	${\Lambda}$	&	${\Sigma}$	&	$\Omega$ \\
\hline
           1 &      0.000000 &       4 &      0.5 &     inf       &  -0.000721 &  + &  e &  X2Pi    &       0 &   1 &     -0.5 &      0.5 \\
           2 &   1970.313482 &       4 &      0.5 &    8.4593E-03 &  -0.000723 &  + &  e &  X2Pi    &       1 &   1 &     -0.5 &      0.5 \\
           3 &   3869.660924 &       4 &      0.5 &    4.5247E-03 &  -0.000724 &  + &  e &  X2Pi    &       2 &   1 &     -0.5 &      0.5 \\
           4 &   5698.780278 &       4 &      0.5 &    3.2392E-03 &  -0.000725 &  + &  e &  X2Pi    &       3 &   1 &     -0.5 &      0.5 \\
           5 &   7458.470416 &       4 &      0.5 &    2.6176E-03 &  -0.000724 &  + &  e &  X2Pi    &       4 &   1 &     -0.5 &      0.5 \\
           6 &   9148.745041 &       4 &      0.5 &    2.2639E-03 &  -0.000723 &  + &  e &  X2Pi    &       5 &   1 &     -0.5 &      0.5 \\
           7 &  10768.764728 &       4 &      0.5 &    2.0464E-03 &  -0.000721 &  + &  e &  X2Pi    &       6 &   1 &     -0.5 &      0.5 \\
           8 &  12316.884393 &       4 &      0.5 &    1.9086E-03 &  -0.000717 &  + &  e &  X2Pi    &       7 &   1 &     -0.5 &      0.5 \\
           9 &  13749.386223 &       4 &      0.5 &    2.4180E+00 &   3.337141 &  + &  e &  a4Sigma &       0 &   0 &      0.5 &      0.5 \\
          10 &  13790.645552 &       4 &      0.5 &    1.8223E-03 &  -0.000713 &  + &  e &  X2Pi    &       8 &   1 &     -0.5 &      0.5 \\
          11 &  15187.015658 &       4 &      0.5 &    1.7723E-03 &  -0.000709 &  + &  e &  X2Pi    &       9 &   1 &     -0.5 &      0.5 \\
          12 &  15859.208556 &       4 &      0.5 &    5.8285E-01 &   3.337143 &  + &  e &  a4Sigma &       1 &   0 &      0.5 &      0.5 \\
          13 &  16502.475508 &       4 &      0.5 &    1.7500E-03 &  -0.000710 &  + &  e &  X2Pi    &      10 &   1 &     -0.5 &      0.5 \\
          14 &  17732.992112 &       4 &      0.5 &    1.7519E-03 &  -0.000707 &  + &  e &  X2Pi    &      11 &   1 &     -0.5 &      0.5 \\
          15 &  17783.585608 &       4 &      0.5 &    2.8270E-01 &   3.337142 &  + &  e &  a4Sigma &       2 &   0 &      0.5 &      0.5 \\
          16 &  18874.096315 &       4 &      0.5 &    1.7780E-03 &  -0.000691 &  + &  e &  X2Pi    &      12 &   1 &     -0.5 &      0.5 \\
          17 &  19508.796071 &       4 &      0.5 &    1.6795E-01 &   3.337143 &  + &  e &  a4Sigma &       3 &   0 &      0.5 &      0.5 \\
          18 &  19921.019747 &       4 &      0.5 &    1.8315E-03 &  -0.000671 &  + &  e &  X2Pi    &      13 &   1 &     -0.5 &      0.5 \\
          19 &  20868.688182 &       4 &      0.5 &    1.9204E-03 &  -0.000643 &  + &  e &  X2Pi    &      14 &   1 &     -0.5 &      0.5 \\
          20 &  21019.734154 &       4 &      0.5 &    1.1238E-01 &   3.337140 &  + &  e &  a4Sigma &       4 &   0 &      0.5 &      0.5 \\
\hline
\hline
\end{tabular}
}
\mbox{}\\
{\flushleft
$n$:   State counting number.     \\
$\tilde{E}$: State energy in \cm. \\
$g_i$:  Total statistical weight, equal to ${g_{\rm ns}(2J + 1)}$.     \\
$J$: Total angular momentum.\\
$\tau$: Lifetime (s$^{-1}$).\\
$g$-Land\'{e} factors. \\
$+/-$:   Total parity. \\
$e/f$:   Rotationless parity. \\
State: Electronic state.\\
$v$:   State vibrational quantum number. \\
$\Lambda$:  Projection of the electronic angular momentum. \\
$\Sigma$:   Projection of the electronic spin. \\
$\Omega$:   Projection of the total angular momentum, $\Omega=\Lambda+\Sigma$. \\
}

\end{table*}

\begin{table}
\caption{Extract from the transitions file of the $^{28}$SiH line list.}
\label{table:trans}
\centering
{\tt
\begin{tabular}{rrrr}
\hline
\multicolumn{1}{c}{$f$}	&	\multicolumn{1}{c}{$i$}	&
\multicolumn{1}{c}{$A_{fi}$ (s$^{-1}$)}	&\multicolumn{1}{c}{$\tilde{\nu}_{fi}$}	
\\
\hline\hline
1494	&	1882	&	6.5269E-07	&	21.092521	\\
9460	&	9670	&	3.4859E-08	&	21.142047	\\
6251	&	6561	&	1.1771E-09	&	21.143927	\\
1582	&	1711	&	1.0066E-10	&	21.144601	\\
575	&	706	&	2.0164E-11	&	21.145126	\\
413	&	290	&	1.4640E-03	&	21.153921	\\
6969	&	7067	&	7.8938E-10	&	21.161879	\\
4731	&	5076	&	1.2289E-11	&	21.165181	\\
2255	&	2642	&	5.3380E-11	&	21.170838	\\
\hline
\end{tabular}
}

\noindent
 $f$: Upper  state counting number;\\
$i$:  Lower  state counting number; \\
$A_{fi}$:  Einstein-A coefficient in s$^{-1}$; \\
$\tilde{\nu}_{fi}$: transition wavenumber in \cm.\\
\end{table}


\section{Examples of spectra}

\begin{figure}
\centering
\includegraphics[width=300pt]{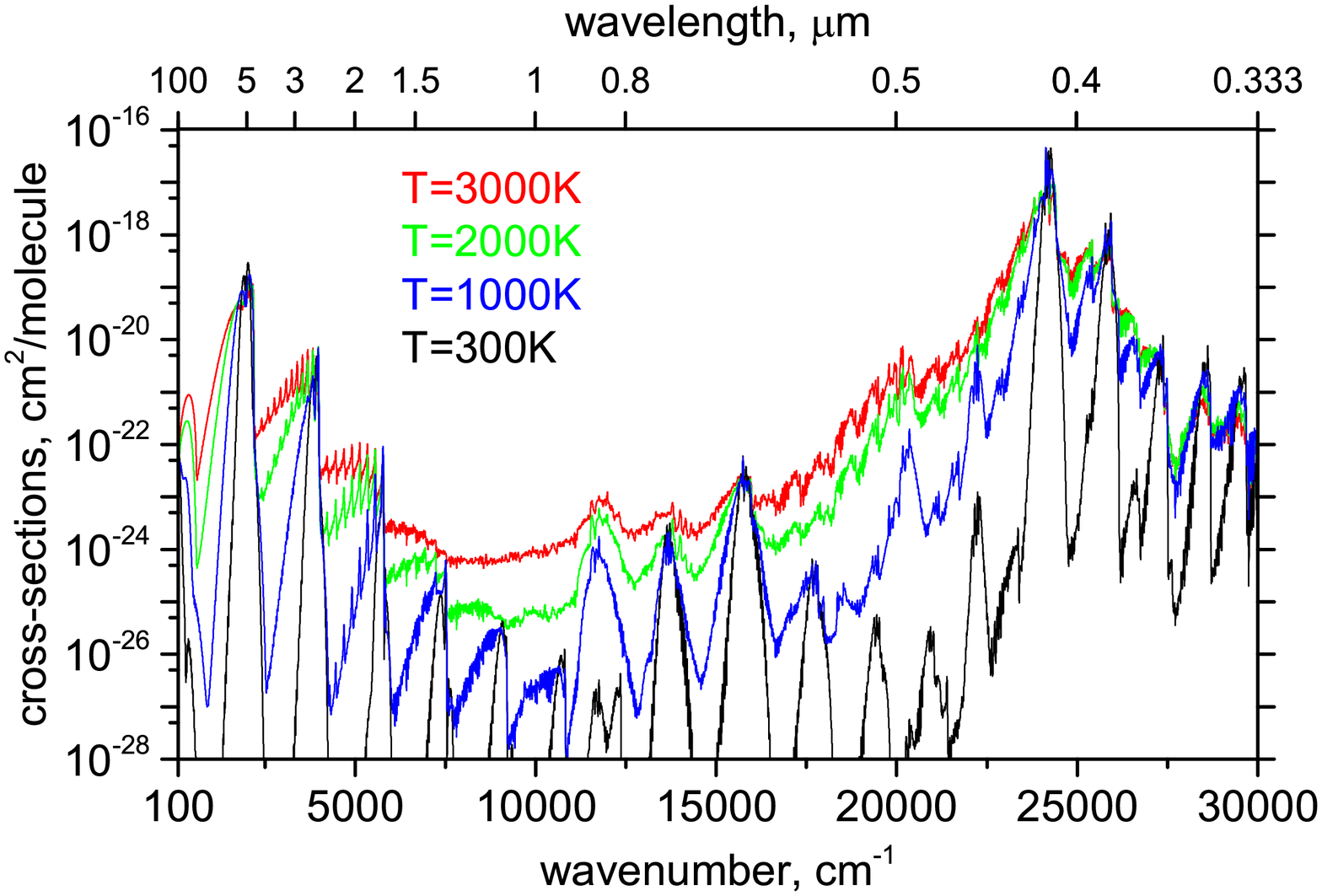}
\caption{Comparison of the SiH absorption spectra at five different temperatures.
The difference in intensity between 300 K and higher temperatures is most
pronounced around 20~000~\cm.}
\label{temp}
\end{figure}

\begin{figure}
\centering
\includegraphics[width=300pt]{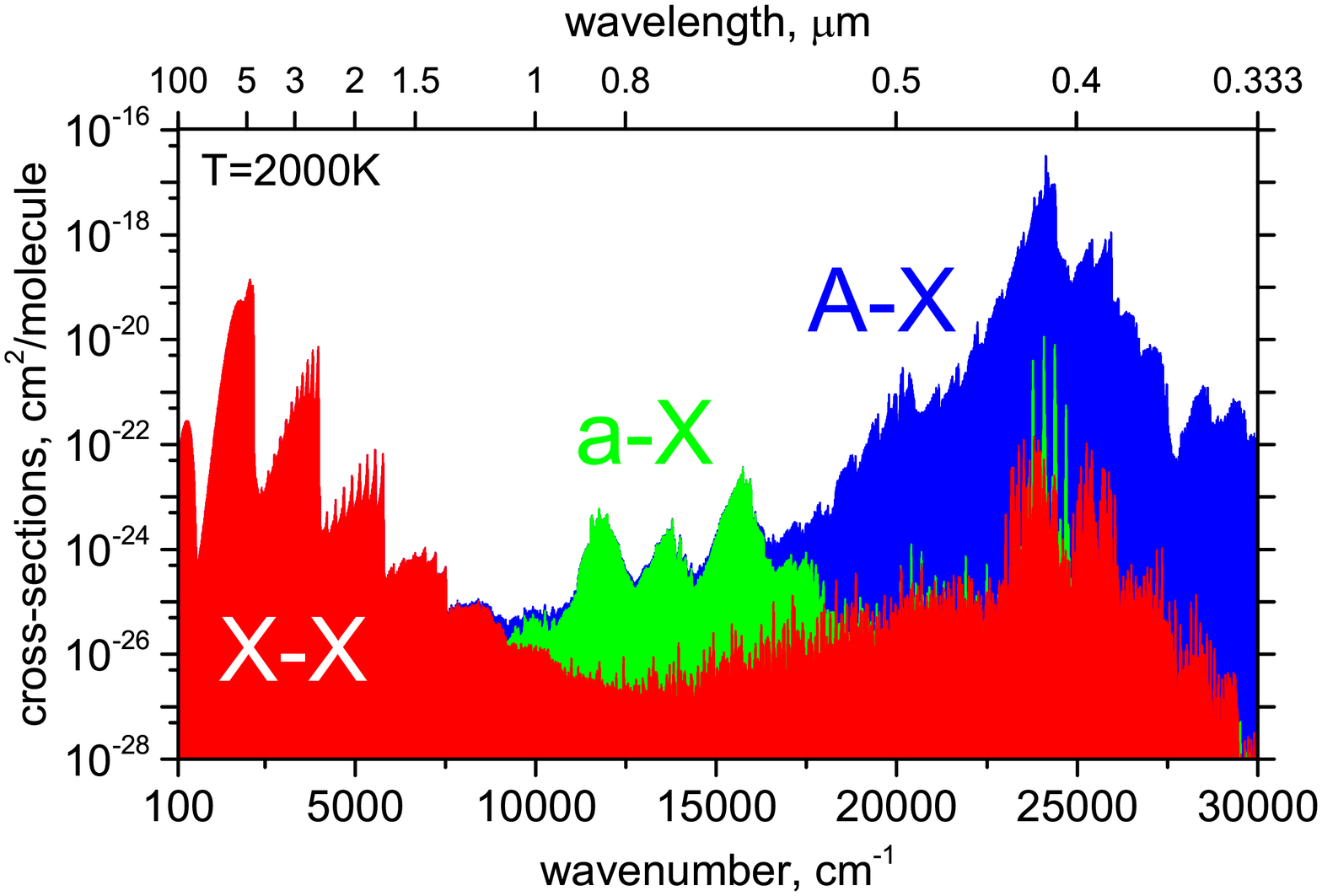}
\caption{$T=2000$~K absorption spectra of $^{28}$SiH: $X$--$X$, $a$--$X$, and $A$--$X$ bands, where
the $a$--$X$ band is dipole forbidden.}
\label{f:bands}
\end{figure}

The temperature at which spectra are simulated has a strong effect on the
intensities produced. Some examples of absorption spectra at different
temperatures are presented in Fig.~\ref{temp}. The structure of the strongest
electronic bands is shown in Fig.~\ref{f:bands}. The  $a$~$^{4}\Sigma^-$ --
$X$~$^{2}\Pi$ band is dipole forbidden but can `steal' intensity by interacting
with the the $X$~$^{2}\Pi$ state. Considering the relative contrast of this band,
it would be interesting to see attempts of detecting this bands experimentally.


\subsection{Comparisons of spectra}

\begin{figure}
\centering
\includegraphics[width=220pt]{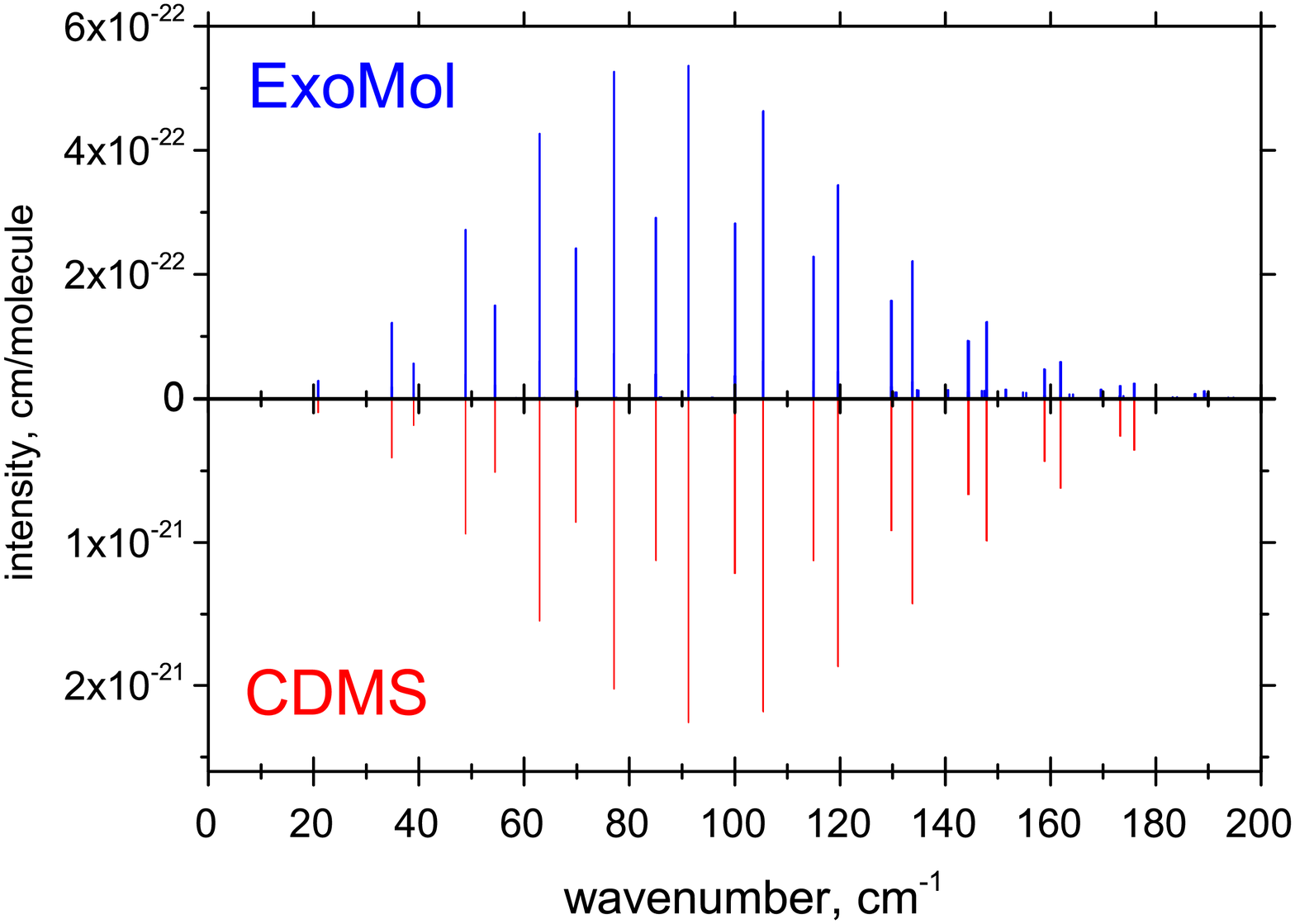}
\caption{Comparison with available CDMS data for the $(0-0)$ transitions of
$^{28}$SiH at 298K. The difference in intensity is about 4.4 times due to the
difference in the dipoles.}
\label{cdms-comp}
\end{figure}

In order to test the quality of our theoretical line list, we present a number
of comparisons with previous works. The CDMS catalogue contains a comparatively
small number of transitions at low wavenumbers \citep{cdms}. For this reason,
this work is only comparable in the $(0-0)$ band.
The CDSM spectra include hyperfine splitting, which is ignored in
our calculations. In order to provide a fair comparison of CDMS with our
spectra, we have combined the CDMS hyperfine sub-structures into single lines.
As can be seen from Fig.~\ref{cdms-comp}, the line positions are very  similar,
but the intensities differ by about a factor of four. The latter is due to the
difference in the dipole moments. The CDMS used the equilibrium \textit{ab
initio} dipole moment value 0.087~D  by \citet{75MeRoxx.SiH}, while our value is
$\mu_{\rm e}$ is 0.097~D. With  our state-of-the-art \ai\ level of theory of
MRCI/aug-cc-pwC5Z-DK we should provide a higher quality $X$--$X$ dipole moment. More importantly, our calculations
also fully incorporate the effect of  zero-point vibrational motion into the dipole matrix element
by using a proper averaging over the vibrational wavefunctions, which is important for this band due to the sign change of the
DMC close to the equilibrium bond length, see Fig.~\ref{f:dipoles}. As the result, for the vibrationally averaged dipole moment $\mu_0$
is 0.0474 D, i.e.  significantly smaller (about half the size) than the equilibrium dipole moment. The
corresponding intensities are therefore 4 times  weaker than those predicted by CDMS. Such an overestimate would lead, for example, to
fourfold under-estimates of the abundance of any observed SiH. We recommend that the CDMS data is updated.

\begin{figure}
\centering
\includegraphics[width=220pt]{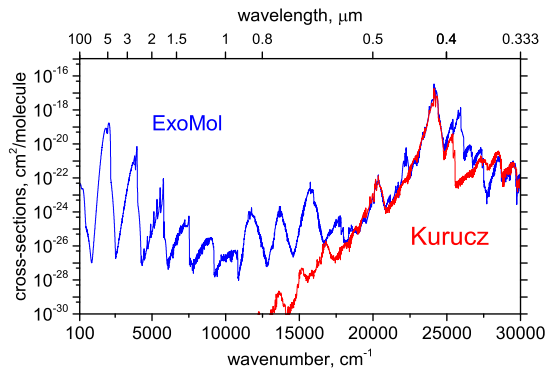}
\caption{A comparison of absorption spectra produced from our line list with that of
\citet{11Kurucz.db} at 1000K. }
\label{kur-comp}
\end{figure}

Figure~\ref{kur-comp} shows a comparison with a spectrum generated using the SiH
line list by \citet{11Kurucz.db}, where only the $A$--$X$ transitions were included. These and  other spectra presented in this work were computed using our new program \textsc{ExoCross} \citep{jt708}. 
In the range where a comparison can be made (above  20~000~\cm), the main peak  at 24~100 \cm\ ($0-0$) agrees quite well, while Kurucz's ($1-0$) and ($0-1$) peaks on the right- and left-hand side from it are much weaker. The ($2-0$) and ($0-2$) bands agree well again, also for the absolute intensities. The disagreement could be due to our use of more modern experimental data, and to the use of higher level \ai\ techniques, causing a discrepancy between the two results.

\begin{figure}
\centering
\includegraphics[width=220pt]{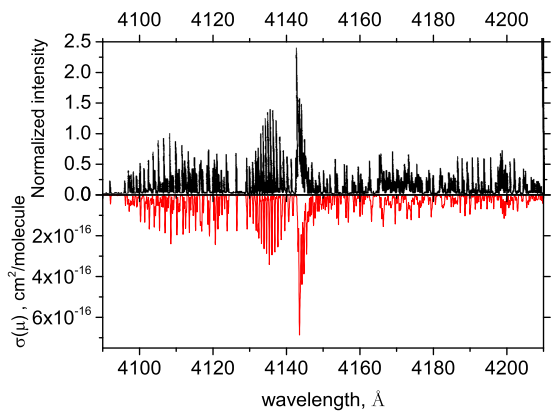}
\caption{Upper display: Experimental emission spectrum of SiH from a 100 mTorr
pure silane radio frequency discharge produced by \citet{97StMaRa.SiH} at 2000~K. Peak
heights are normalised to the highest peak of the R$_1$ branch, R$_1$(10.5).
Lower display: Theoretical emission spectrum of SiH at 2000~K. The peak
intensities are similar to that from the experiment.}
\label{stmara-comp}
\end{figure}

An additional measure of the accuracy of our line list is comparison to
the experimental spectrum produced by \citet{97StMaRa.SiH}, which is presented in
Fig.~\ref{stmara-comp}. The basic shapes of the spectra are the same, with the
peak intensity occurring close to 4140 \AA\, which is taken to be the $Q$
branch. Furthermore, additional peaks at 4130 \AA\ and 4104 \AA\ are replicated,
as is the shape below 4100 \AA. Differences in intensities arise at longer
wavelengths (although the shape remains consistent), most likely because of the
non-thermal effects not properly considered here (we assumed a Boltzmann
distribution and used a Gaussian line profile of 0.5~\cm\ HWHM).

\subsection{Partition Function}

The partition function was calculated with \duo\ in steps of 1~K
and was fitted to the following functional form \citep{jt263} given by
\begin{equation}\label{eq:fit1d}
\log_{10}Q(T) = \sum\limits^{8}_{n=0} a_n (\log_{10}T)^n.
\end{equation}
The fitted expansion parameters for $^{28}$SiH are presented in Table
\ref{table:fit1d}. These parameters reproduce the temperature dependence of
partition function of SiH with a relative root-mean-square error of 0.56~\%;
the fitting error increases to 1.8\%\ at $T$ = 5000~K. This is still a very
small error, and thus the fit can be said to reliably reproduce the partition
function. The partition function and the expansion parameters for all four
species are included into the supplementary materials.

\begin{table}
\centering
\caption{Expansion coefficients for the partition function of $^{28}$SiH given
by
Eq.~(\ref{eq:fit1d}). Parameters for other isotopologues can be found in the
supplementary material.}
 \begin{tabular}{cc}
 \hline\hline
$a_i$ &  value	\\ \hline
$a_0    $&  0.872519166079  \\
$a_1    $&  0.331757300487  \\
$a_2    $&  -1.370132158260 \\
$a_3    $&  2.297149082100  \\
$a_4    $&  -1.997501906630 \\
$a_5    $&  1.134338070560  \\
$a_6    $&  -0.384224515822 \\
$a_7    $&  0.068401930630  \\
$a_8    $&  -0.004881419966 \\
\hline
\end{tabular}
\label{table:fit1d}
\end{table}

\begin{figure}
\centering
\includegraphics[width=220pt]{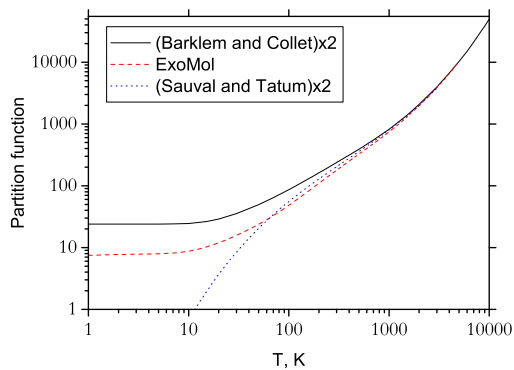}
\caption{A comparison between the partition function produced from our line list
and the theoretical partition function described by \citet{84SaTaxx.partfunc}.
The latter values were multiplied by $g_{\rm ns}$ =2.}
\label{part-function}
\end{figure}

Our partition function can be compared to that computed by
\citet{84SaTaxx.partfunc} and by \citet{16BaCoxx.partfunc}. In order to directly
compare with our calculated partition function, these partition functions need
to be multiplied by the nuclear statistical weight $g_{ns}=2$ \citep{jt631}. This is  the so-called  physics convention for the nuclear statistical weights, which ExoMol uses. Figure \ref{part-function} shows this comparison.
All three partition function agree well at high temperature.
The values given by \citet{16BaCoxx.partfunc} at lower temperatures are too high.
We believe that our
function is sufficiently complete and more accurate at lower temperatures.
 At
$T=75$~K  our partition function 35.278 compares well to the  CDMS's pure
rotational value at $T=75$~K of 35.277; this agreement continues to very low
temperature, where our partition function has the physically correct behaviour at $T \rightarrow 0$~K:
\begin{equation}
Q(T) \approx 4 (e^{-0\, c_2/T} +  e^{-0.097\, c_2/T} ) + \ldots .
\end{equation}
where 4 is the total degeneracy, 0 and 0.097~\cm\ are the two lowest term values at $J=0.5$, and $c_2$ is the second radiation constant.

The nuclear statistical weights $g_{\rm ns}$ used to produce the partition functions for $^{28}$SiD, $^{29}$SiH, and $^{30}$SiH (as well as their line lists) are 3, 4 and 2, respectively.

\section{Conclusion}

Accurate and complete line lists for $^{28}$SiH and three minor isotopologues
$^{29}$SiH, $^{30}$SiH and $^{28}$SiD have been produced,
displaying good agreement with existing theoretical and experimental data.
The accuracy of the rotational line positions is comparable to the one obtainable with
effective rotational Hamiltonians. In order to reproduce the $\Lambda$-doubling splitting,
an interaction via an electronic angular momentum coupling curve
with the $B$~$^{2}\Sigma^-$ state was included using a very simple approximation to represent the PEC of the $B$-state.

The vibrationally averaged dipole moment $\mu_0$ in the ground electronic state exhibit strong vibrational
dependence and is about half in magnitude with respect to the
equilibrium value. We suggest that the CDMS value of $\mu_{0}$ should be updated.

The lifetimes computed for the $A$--$X$ rotational transitions ($J\le 13.5$) using our best \ai\ transition DMC ($\approx 400$~ns) are off by about a factor 1.3 from the corresponding experimental values ($530$~ns). We use this experimental value to improve the \ai\ transition dipole moment curve
by scaling it by a factor $\sqrt{400/530} \approx 0.869$, which leads to  identical lifetimes (within 0.1~ns).
A higher level of \ai\ theory is therefore needed in order to be confident
that the dipole moments used are accurate. It would also be beneficial to have more
experimental data sensitive to the dipole moment, e.g. intensities or lifetimes.

\begin{figure}
\includegraphics[width=0.5\textwidth]{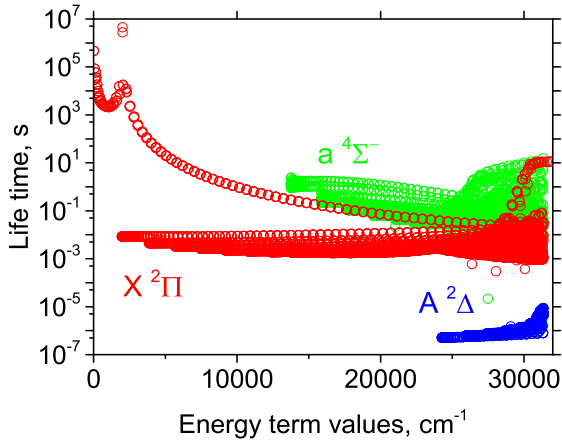}
\caption{Lifetimes of the three lower electronic states of SiH.}
\label{f:lifetime}

\end{figure}

\section*{Acknowledgements}

This work was supported by the UK Science and Technology Research Council (STFC)
No. ST/M001334/1 and the COST action MOLIM No. CM1405.  This work made extensive
use of UCL's Legion  high performance computing facility.

\bibliographystyle{mnras}

\label{lastpage}
\end{document}